\documentclass[showpacs,twocolumn,prb]{revtex4}
\usepackage{graphicx}
\newcommand{\be}{\begin{equation}}
\newcommand{\ee}{\end{equation}}
\newcommand{\bea}{\begin{eqnarray}}
\newcommand{\eea}{\end{eqnarray}}
\newcommand{\bwt}{\begin{widetext}}
\newcommand{\ewt}{\end{widetext}}

\begin{document}
\title{Optical Sum Rule anomalies in the High-T$_c$ Cuprates}
\author{F. Marsiglio}
\affiliation{Department of Physics, University of Alberta, Edmonton, Alberta,
Canada, T6G~2G7}
\begin{abstract}
We provide a brief summary of the observed sum rule anomalies in the high-T$_c$ cuprate materials.
A recent issue has been the impact of a non-infinite frequency cutoff in the experiment. In the
normal state, the observed anomalously high temperature dependence can be explained as a `cutoff effect'. The anomalous rise in the optical spectral weight below the superconducting transition,
however, remains as a solid experimental observation, even with the use of a cutoff frequency. 
\end{abstract}

\date{\today}
\maketitle

\section{background}
There is a global optical sum rule, due to Kubo, \cite{kubo57} that
states that the sum total of the optical spectral weight is a constant of
the material, and proportional to the electron density in that material.
An excellent demonstration of this sum rule was provided for Aluminum, \cite{smith78}
where an energy range on the order of hundreds of keV's was required before all
13 electrons could be accounted for.

In contrast, in the single band optical sum rule, \cite{kubo57} the total
optical spectral weight of a single band is a weighted average of
the second partial derivative of the single particle energy over the
occupied states of that band in the first Brillouin. This sum rule is only
useful experimentally if a reasonable energy separation exists between intraband transitions
involving the band in question, and all other transitions involving the other bands in the
material.

The usual philosophy in condensed matter physics is to focus on low energy scales. For a
metal this means that we are interested in states near the Fermi energy. When some phenomenon
(say, a transition) occurs at low temperature, then the temperature often sets the scale for
what is meant by `low'. So it is that the Ferrell-Glover-Tinkham sum rule, \cite{ferrell58},
which is a special case of the global Kubo sum rule applied to the superconducting transition,
can often be checked by measuring the changes in the optical absorption up to $\omega \approx
20-30T_c$ \cite{ferrell58} (as compared to the more challenging $\omega \rightarrow \infty$).

Thus the prediction by Hirsch, \cite{hirsch92} of an apparent sum rule violation in the cuprates,
came both as a surprise and as an unlikely possibility. His suggestion was motivated by the
possibility of superconductivity driven by kinetic energy lowering (compared with the
paradigm provided by the BCS theory of conventional, potential energy-driven superconductivity).
These ideas were motivated by the so-called hole mechanism of superconductivity that
had originated a few years earlier.\cite{hirsch89,marsiglio90} Around the same time Anderson
and coworkers \cite{wheatley88} advanced similar ideas, but with specifically interplanar kinetic
energy lowering. This was known as the Inter-layer Tunneling (ILT) Theory.

\section{the experiments}

Seven years after Hirsch's prediction, Basov and coworkers \cite{basov99} measured an apparent
c-axis sum rule violation in Tl$_2$Ba$_2$CuO$_{6+x}$ (Tl2201). Around this time, however, the
ILT theory became untenable because it could not account for the condensation energy.
\cite{loram94,schutzmann97,moler98} Theoretical work at this time \cite{hirsch00}
clarified that the sum rule violation should occur in the in-plane response.
This was followed by a number of experiments that probed the in-plane optical sum rule, as a function of temperature, for a variety of cuprate materials. For example, in Refs.
\onlinecite{molegraaf02,santander-syro03,ortolani05,carbone06b,vanheumen07}, anomalies were
found in both/either the normal state temperature dependence and and/or the transition
into the superconducting state. One group \cite{boris04} remains unconvinced that any anomalous behaviour exists in YBa$_2$Cu$_3$O$_{6.9}$, although controversy still surrounds this issue.
\cite{kuzmenko05,santander-syro05} 

Before we summarize the anomalous behavior observed, we first write down Kubo's single
band sum rule. It is 
\begin{equation}
W(T) = {2 \hbar^2 \over \pi e^2}\int_{0}^{+\infty }d\nu \mathop{\rm
Re} \left[ \sigma_{\rm xx} (\nu )\right] = \frac{2}{N}\sum_{k}
{\partial^2\epsilon_k \over \partial k_x^2} n_{k},%
\label{sumrule}
\end{equation}
where the left hand side is proportional to the integrated conductivity (real part)
arising solely from intraband transitions. This conductivity has its strongest intensity
at low frequencies; nonetheless the integration extends to infinite frequency, where
`long tail' contributions can be significant. In practice experimenters must cut off this 
integration in their data at some finite frequency (typically 1-2 eV), where they believe
most of the contribution has been accounted for, and, at the same time, interband contributions
have not been included. The right-hand-side (RHS) of this equation involves only single particle
properties, namely the second derivative of the dispersion relation, and the single particle
occupation number, $n_k$. The RHS has been identified with the kinetic energy associated with the
single band. This is because in the special case of a nearest neighbour tight-binding model,
the RHS is proportional to the kinetic energy $<K>$,
\begin{equation}
<K> =  \frac{2}{N}\sum_{k} \epsilon_k  n_{k}.%
\label{kinetic}
\end{equation}
Thus, specifically in the case of nearest neighbour hopping only, we
have, in two dimensions \cite{hirsch00},

\begin{equation}
W(T) = -<K>/2.%
\label{relation}
\end{equation}

Note that for parabolic bands, the single band sum rule looks just like the global optical
sum rule, except that the electron density pertains only to that contained within the single band.
Obviously, for more realistic band structures, the identification of the sum of the optical
spectral weight $W$ with the kinetic energy $<K>$ will not be precise, but it is qualitatively
similar for a variety of scenarios. {\em However}, a thorough investigation was carried out in
Ref. \onlinecite{marsiglio06b}; a {\em qualitative} distinction between the two properties will
occur near a van Hove singularity. For the cases studied in Ref. \onlinecite{marsiglio06b} this
possibility did not appear to be relevant for the experiments performed up to that time, but
this distinction should be kept in mind. Nonetheless, for the rest of this review, we will assume 
that the kinetic energy can be identified with the optical spectral weight.

In Fig.~1 we show a {\em schematic} plot of the expected temperature dependence for the single
band optical spectral weight. In the normal state the expected $W(T)$ follows a $T^2$ dependence, which can be derived from a Sommerfeld expansion.\cite{benfatto06} At $T_c$, because kinetic energy is sacrificed in the usual BCS transition, the expected spectral weight {\em decreases} as the temperature is lowered, so a marked kink occurs at $T_c$.
\begin{figure}[tp]
\begin{center}
\includegraphics[height=3.8in,width=3.4in]{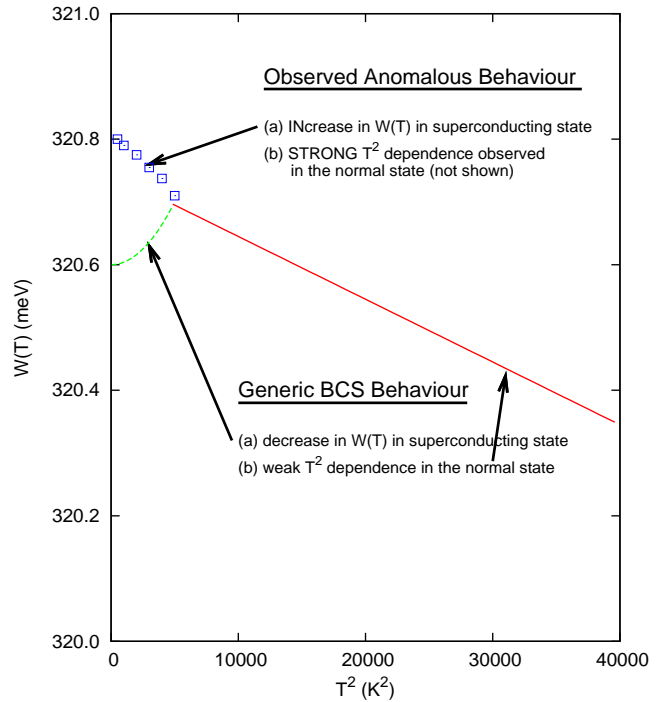}
\caption{The expected temperature dependence for the optical sum rule, given by
$W(T)$; in the normal state $W(T)$ is expected to be proportional to $T^2$, with a
small proportionality constant indicative of the smallness of the ratio $T/D$, where $D$
is the electronic bandwidth. This result is plotted with the solid (red) curve. In the superconducting state a decrease is expected, as indicated by the dashed (green) curve, thus
forming a kink at $T_c$. The observed normal state generally has a much higher slope, and,
at least for underdoped and some optimally doped samples, the optical sum rule exhibits an 
{\em increase} below the superconducting transition temperature (indicated by the blue symbols).}
\end{center}
\end{figure}
In contrast the observed normal state temperature dependence is much stronger than expected.
Various groups have inferred this to be indicative of correlations and/or phase fluctuations
(see, for example, Refs. \onlinecite{haule07,toschi05,eckl03}). A much simpler explanation was
suggested almost immediately, \cite{karakozov02} and was supported by more recent numerical and analytical work. \cite{norman07,marsiglio08}. This will be further discussed below.

The superconducting state anomaly can be viewed as an indication of scattering rate collapse
\cite{norman02,knigavko04,schachinger05,knigavko05};
a phenomenology \cite{marsiglio06} developed by the present author shows this most clearly, and
relates the present experiment to a similar phenomenon first seen in microwave experiments.
\cite{nuss91,bonn92,hosseini99,harris06}. As far as we are concerned, this phenomenology adequately
`explains' the anomaly seen to occur below the superconducting transition temperature. Of course
microscopic models continue to be proposed and vie for the goal of explaining superconductivity in
the cuprates at a microscopic level. In the remaining part of this report we review recent developments concerning the challenge \cite{karakozov02} that all the anomalous observations could be due to a `cutoff effect', i.e. the non-infinite frequency cutoff used by experimentalists on the left-hand-side (LHS) of  Eq. (\ref{sumrule}) could be the cause of the anomalous temperature dependence, both in the normal and superconducting states.

\section{some theory}

The resolution of this possibility requires, for some of the various theoretical models studied so far, an accurate evaluation of the LHS of Eq. (\ref{sumrule}), specifically when the integration is {\em not} taken to infinite frequency. This was done recently in the normal state by Norman et al. \cite{norman07} for two models where electrons interact with a boson. In both cases they determined the optical conductivity, and evaluated 
\begin{equation}
\tilde{W}_{\rm app}(\Omega_C) = {2 \hbar^2 \over \pi e^2}\int_{0}^{\Omega_c}d\nu \mathop{\rm
Re} \left[ \sigma_{\rm xx} (\nu )\right],
\label{app_sumrule}
\end{equation}
which gives approximately the sum rule required in Eq.  (\ref{sumrule}). In this way they could explore whether or not the strong temperature dependence in the normal state could be induced artificially by the non-infinite cutoff (as experimentalists are required to utilize). They indeed found that "...the bulk of the observed $T$ dependence in the normal state is related to the finite cutoff". This is consistent with the original suggestion (in the context of electron-phonon scattering) of Karakozov et al. \cite{karakozov02}.

Karakozov et al. \cite{karakozov02} also suggested, based on the Mattis-Bardeen formula (dirty limit) for the conductivity in the superconducting state, that an anomalous temperature dependence below the superconducting transition temperature may also be due to a finite cutoff, as used in Eq. (\ref{app_sumrule}). We therefore investigated this possibility in Ref. \onlinecite{marsiglio08}.

In fact, for the Mattis-Bardeen limit, our numerical work supports the result of Karakozov
et al. However, a more realistic procedure is to utilize a non-infinite scattering rate (see
Refs. \onlinecite{marsiglio97,marsiglio08_chap} for pertinent formulas). A key result, derived to be accurate at high frequency, for any value of scattering $1/\tau$, is
\bea
{\sigma_{1S}(\nu) \over \sigma_0} \approx &&{ (1/\tau)^2 \over \nu^2 + (1/\tau)^2} \biggl(
1 - 2 \bigl({\Delta_0 \over \nu}\bigr)^2 \bigl[1 + \log{2 \nu \over \Delta_0} \bigr] \nonumber \\
&& \phantom{aaa} - 2 { \Delta_0^2 \over \nu^2 + (1/\tau)^2} \bigl[1 - 2\log{2 \nu \over \Delta_0} \bigr] \biggr).
\label{approx1}
\eea
\begin{figure}[tp]
\begin{center}
\includegraphics[height=3.2in,width=3.0in, angle = -90]{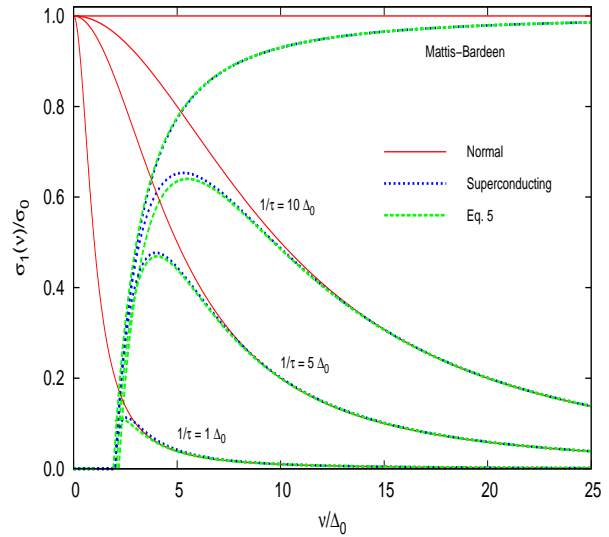}
\caption{(color online) The real part of the optical conductivity (normalized to
$\sigma_0 \equiv {\omega_p^2 \over 4 \pi} \tau$)
vs. frequency, for a variety of scattering rates. The normal state is shown by the solid (red) curves, for the scattering rates indicated.
The numerical results in the superconducting state are given by the dotted (blue) curves. The dashed (green) curves indicate the high frequency expansions for the superconducting state (Eq. (\protect\ref{approx1})). These are surprisingly very accurate right down to $\nu = 2 \Delta_0$. Except for the Mattis-Bardeen limit, the superconducting state crosses
the normal state (not discernable on this figure), a characteristic first noted by Chubukov et al. \protect\cite{chubukov03}. This implies that the sum rule correction will be opposite to that in the Mattis-Bardeen limit, i.e. in the direction opposite to that observed in experiment.}
\end{center}
\end{figure}
This is shown in Fig.~2, for a variety of scattering rates, along with the Mattis-Bardeen limit.
A first observation is that the approximation Eq. (\ref{approx1}) is remarkably accurate for 
practically all frequencies down to $2\Delta_0$. We suggest that Eq. (\ref{approx1}) will serve as
a simple way to characterize otherwise complicated calculations.\cite{marsiglio97,marsiglio08_chap}.
A second observation, not really visible in Fig.~2, but apparent in an expanded version (see Figs.6 and 7 in Ref. \onlinecite{marsiglio08}), is that the correction coming from integrating the results in Fig.~2 up to some non-infinite cutoff (as required in the approximate optical sum rule, Eq. (\ref{app_sumrule})) will be opposite to the anomalous behaviour observed in experiment. This means that the observed anomaly below the superconducting transition {\em cannot} be explained by a non-infinite frequency cutoff used in the experimental procedure.

\section{summary}

A number of experimental groups have observed anomalies in underdoped samples of various high $T_c$ materials. This has been interpreted as signalling new physics. More specifically, in the normal state it appears to indicate strong correlations, or phase fluctuations, or both. In the superconducting state the anomaly indicates a new kind of 'kinetic energy-driven' superconductivity. Model calculations taking account of a non-infinite frequency cutoff to evaluate the sumrule indicate that the observations in the normal state may be explained by the use of such a cutoff, i.e. they are an artifact. On the other hand, the use os such a cutoff {\em cannot} explain the anomaly that occurs below the superconducting transition temperature, and the original interpretation of this anomaly as a signature for a kinetic energy-driven mechanism remains viable.

\begin{acknowledgments}

This work was supported in part by the
Natural Sciences and Engineering Research Council of Canada (NSERC),
by ICORE (Alberta), and by the Canadian Institute for Advanced Research
(CIfAR).  I am grateful to the Aspen Center for
Physics, where some of the ideas for this work were first solidified; discussions with
Mike Norman and Andrey Chubukov, both in Aspen and earlier in Rome at the Optical Sum Rules workshop were most helpful. Finally, I wish to thank the many collaborators whose work I very briefly reviewed here.
\end{acknowledgments}

\bibliographystyle{prb}

\end{document}